\documentclass[prl,twocolumn,groupedaddress,superscriptaddress]{revtex4-1}
\usepackage{amsmath}
\usepackage{graphicx,color}

\usepackage[utf8]{inputenc}
\usepackage{siunitx}

\newcommand{\PRLsection}[1]{\noindent {\it#1} -}
\newcommand{\var}{{\rm var}}
\newcommand{\expect}[1]{\left<#1\right>}
\newcommand{\bea}{\begin{eqnarray}}
\newcommand{\eea}{\end{eqnarray}}
\newcommand{\nnp}{\nonumber \\ & & +}
\newcommand{\ket}[1]{\left|#1\right>}
\newcommand{\bra}[1]{\left<#1\right|}
\newcommand{\muzero}{{{\mu}}}

\newcommand{\NPulses}{N_{R}}

\begin{document}

\title{Efficient quantification of non-Gaussian spin distributions}

\author{B. Dubost}
\email[]{brice.dubost@icfo.es}
\affiliation{ICFO-Institut de Ciencies Fotoniques, Mediterranean Technology Park, 08860 Castelldefels (Barcelona), Spain}
\affiliation{Univ Paris Diderot, Sorbonne Paris Cité, Laboratoire Matériaux et Phénomènes Quantiques, UMR 7162, Bât. Condorcet, 75205 Paris Cedex 13, France}
\author{M. Koschorreck}
\affiliation{Cavendish Laboratory, University of Cambridge, JJ Thompson Avenue, Cambridge CB3 0HE, United Kingdom.}
\author{M. Napolitano}
\author{N. Behbood}
\author{R.J. Sewell}
\affiliation{ICFO-Institut de Ciencies Fotoniques, Mediterranean Technology Park, 08860 Castelldefels (Barcelona), Spain}
\author{M.W. Mitchell}
\affiliation{ICFO-Institut de Ciencies Fotoniques, Mediterranean Technology Park, 08860 Castelldefels (Barcelona), Spain}
\affiliation{ICREA-Instituci\'{o} Catalana de Recerca i Estudis Avan\c{c}ats, 08015 Barcelona, Spain}

\begin{abstract}
We study theoretically and experimentally the quantification of non-Gaussian distributions via non-destructive measurements.  Using the theory of cumulants, their unbiased estimators, and the uncertainties of these estimators, we describe a quantification which is simultaneously efficient, unbiased by measurement noise, and suitable for hypothesis tests, e.g., to detect non-classical states.  The theory is applied to cold $^{87}$Rb spin ensembles prepared in non-gaussian states by optical pumping and measured by non-destructive Faraday rotation probing.  We find an optimal use of measurement resources under realistic conditions,  e.g., in atomic ensemble quantum memories. 

\end{abstract}

\pacs{42.50.Dv,42.50.Lc,03.65.Wj}

\maketitle

\PRLsection{Introduction}
Non-Gaussian states are an essential requirement for universal quantum computation \cite{PhysRevA.68.042319,PhysRevLett.82.1784} and several quantum communication tasks with continuous variables, including improving the fidelity of quantum teleportation \cite{PhysRevA.76.022301} and entanglement distillation \cite{PhysRevLett.89.137903,PhysRevA.66.032316}.
Optical non-Gaussian states have been demonstrated \cite{PhysRevLett.97.083604,Ourjoumtsev2007,Wakui:07,PhysRevLett.101.233605,PhysRevLett.107.213602} and  proposals in atomic systems \cite{PhysRevLett.91.060401,PhysRevA.79.023841,PhysRevA.79.043808,0295-5075-83-6-60004} are being actively pursued. In photonic systems, histograms \cite{PhysRevLett.92.153601} and 
state tomography  \cite{PhysRevLett.107.213602,PhysRevLett.97.083604,Ourjoumtsev2007,PhysRevLett.101.233605} have been used to show non-Gaussianity, but require a large number of measurements.  For material systems with longer time-scales these approaches may be prohibitively expensive. Here we demonstrate the use of cumulants, global measures of distribution shape, to show non-Gaussianity in an atomic spin ensemble. Cumulants can be used to show non-classicality \cite{PhysRevA.83.052113,PhysRevA.71.011802,arxiv1110.3060}, can be estimated with few measurements and have known uncertainties, a critical requirement for proofs of non-classicality.

\PRLsection{Approach}
Quantification or testing of distributions has features not encountered in quantification of observables. For example, experimental measurement noise appears as a distortion of the distribution that cannot be ``averaged away'' by additional measurements. As will be discussed later, the theory of cumulants naturally handles this situation. We focus  on the fourth-order cumulant $\kappa_4$, the lowest-order indicator of non-Gaussianity in symmetric distributions such as Fock \cite{PhysRevLett.87.050402} and ``Schr\"odinger kitten'' states \cite{Ourjoumtsev2007,PhysRevLett.91.060401}.  We study theoretically and experimentally the noise properties of Fisher's unbiased estimator of $\kappa_4$, i.e., the fourth ``k-statistic'' $k_4$, and find optimal measurement conditions.  Because $\kappa_4$ is related to the negativity of the Wigner function \cite{PhysRevA.83.052113}, this estimation is of direct relevance to detection of non-classical states.  We employ quantum non-demolition measurement, a key technique for generation and measurement of non-classical states in atomic spin ensembles \cite{Appel07072009, PhysRevLett.105.093602} and nano-mechanical oscillators \cite{HertzbergNP2010}.

\PRLsection{Moments, cumulants and estimators}
A continuous random variable $X$ with probability distribution function
$P(X)$ is completely characterized by its moments $\muzero_k \equiv
\int X^k P(X)dX $ or cumulants $ {\kappa}_{n} = \mu_n-\sum_{k=1}^{n-1} \left( ^{
n-1}_{k-1}\right)  \mu_{n-k}{\kappa}_{k}$, where $\left( ^{n}_{k}\right)$ is the binomial coefficient.

Since Gaussian distributions have $\kappa_{n>2}=0$, estimation of $\kappa_4$, (or $\kappa_3$ for non-symmetric distributions), is a natural test for non-Gaussianity.  In an experiment, a finite sample $\{X_1 \ldots X_N\}$ from $P$ is used to estimate the $\kappa$'s.  Fisher's unbiased estimators, known as ``k-statistics'' $k_n$, give the correct expectation values $\expect{k_n} = \kappa_n$  for finite $N$ \cite{AdvTheoStat1958}.   Defining $S_n = \sum_i X_i^n$ 
we have:
\begin{eqnarray}
\label{eq:k3}
{k}_3 & = & ({2 S_1^3 - 3 N S_1 S_2 + N^2 S_3})/N_{(2)} \\
\label{eq:k4}
{k}_4 & = & \left(-6 S_1^4 + 12 N S_1^2 S_2 - 3 N(N-1) S_2^2 
\right. \nonumber \\ & & \left.  - 4 N(N-1) S_1 S_3
+ N^2(N+1) S_4 \right)/N_{(3)}
\end{eqnarray}
where $N_{(m)} \equiv N (N-1) \ldots (N-m)$ .

\begin{figure}
\includegraphics[width=\linewidth]{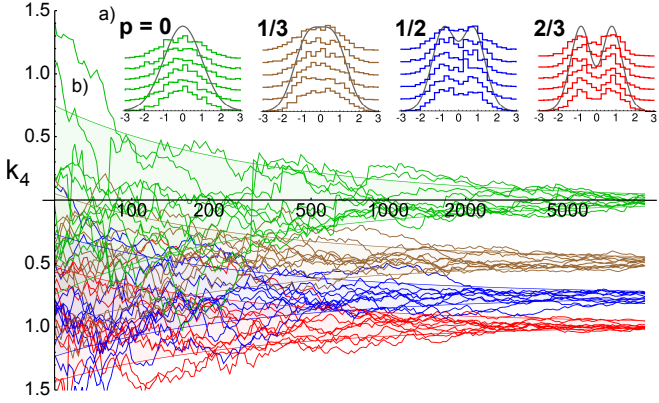}%
\caption{(color online) Simulated estimator $k_4$ as a function of sample size $N$. a) (insets)  black curves show quadrature distributions of states $\rho = (1-p) \ket{0}\bra{0} + p \ket{1}\bra{1}$, scaled to unit variance, and six $N=1000$ histograms (offset for clarity) for $p=0$ (green), $1/3$ (brown), $1/2$ (blue) and $2/3$ (red).  b)  Ten realizations of $k_4$ versus $N$ drawn from each of the four distributions.  Shaded regions show $\kappa_4 \pm \sqrt{\var (k_4)}$, from Eqs (\ref{eq:k4}), (\ref{eq:vark4}). With $N=1000$, $k_4$ distinguishes $p=1/2$ (blue) from $p=0$ (green, Gaussian) with $>7\sigma$ significance, even though the histograms look  similar ``to the eye.'' 
\label{fig:Evolution}}
\end{figure}

We need the uncertainty in the cumulant estimation to test for non-Gaussianity, or to compare non-Gaussianity between distributions.   For hypothesis testing and maximum-likelihood approaches, we need  the variances of $k_3, k_4$ for a given 
$P$. These are found by combinatorial methods and given in reference \cite{AdvTheoStat1958}:
\begin{eqnarray}
\label{eq:vark3}
\var(k_3) & =& {{\kappa}_{6}}/{N} + 9N ({\kappa}_{2}{\kappa}_{4}+ {\kappa}_{3}^2)/N_{(1)} +  6N^2 {\kappa}_{2}^3/N_{(2)} \;\;\;\;\; \\
\label{eq:vark4}
\var(k_4) & =& {{\kappa}_{8}}/{N} + 2 N({8 {\kappa}_{6}{\kappa}_{2} + 24 {\kappa}_{5}{\kappa}_{3} + 17 {\kappa}_{4}^2})/N_{(1)} \nonumber \\
		    &&+ 72 N^2 ({{\kappa}_{4}{\kappa}_{2}^2 + 2 {\kappa}_{3}^2 {\kappa}_{2}})/N_{(2)}   \nonumber \\
	& 	    &+ {24 N^2 (N+1) {\kappa}_{2}^4}/N_{(3)} .
\end{eqnarray}
 It is also possible to estimate the uncertainty in $k_4$ from data $\{X\}$ using estimators of higher order cumulants \cite{AdvTheoStat1958}. The efficiency of cumulant estimation is illustrated in Fig. \ref{fig:Evolution}.

\PRLsection{Measurement noise}
When the measured signal is $Z=X+Y$, where $X$ is the true value and $Y$ is uncorrelated noise, the measured distribution is the convolution $P(Z) = P(X) \otimes P(Y)$.  The effect of this distortion on cumulants is the following: for independent variables, cumulants accumulate (i.e., add) \cite{AdvTheoStat1958}, so that  $\kappa_{n}^{(Z)} = \kappa_{n}^{(X)} +\kappa_{n}^{(Y)}$, where $\kappa_{n}^{(Q)}, k_{n}^{(Q)}$ indicate $\kappa_n, k_n$ for distribution $P(Q)$.  The extremely important case of uncorrelated, zero-mean Gaussian noise, $\kappa_2^{(Y)} = \sigma^2_Y$ and other cumulants zero, is thus very simple: $\kappa_{n}^{(Z)} = \kappa_{n}^{(X)}$ except for $\kappa_{2}^{(Z)} = \kappa_{2}^{(X)} +\sigma^2_Y$.  Critically, added Gaussian noise does not alter the observed $\kappa_3$, $\kappa_4$.

\PRLsection{Experimental system and state preparation} 
We test this approach by estimating non-Gaussian spin distributions in an atomic ensemble, similar to ensemble systems being developed for quantum networking with non-Gaussian states \cite{JHMPersComm}.  The collective spin component $F_z$ is measured by Faraday rotation using optical pulses.   The detected Stokes operator is  $S_y^{(\rm out)} = S_y^{(\rm in)} + G N_{\rm L} F_z/2 $, where $G$ is a coupling constant, $N_{\rm L} $ is  the number of photons,  and $S_y^{(\rm in)}$ is the input Stokes operator, which contributes quantum noise.  In the above formulation $X=  F_z,$ $Y= 2 S_y^{(\rm in)}/(G N_{\rm L})$ and $Z = 2 S_y^{(\rm out)}/(G N_{\rm L})$.

The experimental system is described in detail in references \cite{PhysRevA.79.043815,PhysRevLett.104.093602,PhysRevLett.105.093602}. An ensemble of $\sim10^6~^{87}$Rb atoms is trapped in an elongated dipole trap made from a weakly focused \SI{1030}{nm} beam and cooled to \SI{25}{\micro K}.  
A non-destructive measurement of the atomic state is made using pulses of linearly polarized light detuned \SI{800}{MHz} to the red of the $F=1 \,\rightarrow \,F'=0$ transition of the D$_2$ line and sent through the atoms in a beam matched to the transverse cloud size.  The pulses are of \SI{1}{\micro s} duration, contain $3.7\times 10^6$  photons on average, and are spaced by \SI{10}{\micro s} to allow individual detection.  The 240:1 aspect ratio of the atomic cloud creates a strong paramagnetic Faraday interaction 

$G \approx {6 \times 10^{-8}}$ {rad}/spin.  After interaction with the atoms, $S_y^{(\rm out)}$  is detected with a shot noise limited (SNL) balanced polarimeter in the $\pm 45^\circ$ basis.  $N_{\rm L}$ is measured with a beam-splitter and reference detector before the atoms.  The probing-plus-detection system is shot-noise-limited above $3 \times 10^5$ photons/pulse.   Previous work with this system has demonstrated QND measurement of the collective spin $F_z$ with an uncertainty of $\sim 500$ spins \cite{PhysRevLett.104.093602,PhysRevLett.105.093602}.  

\begin{figure}
\includegraphics[width=\linewidth]{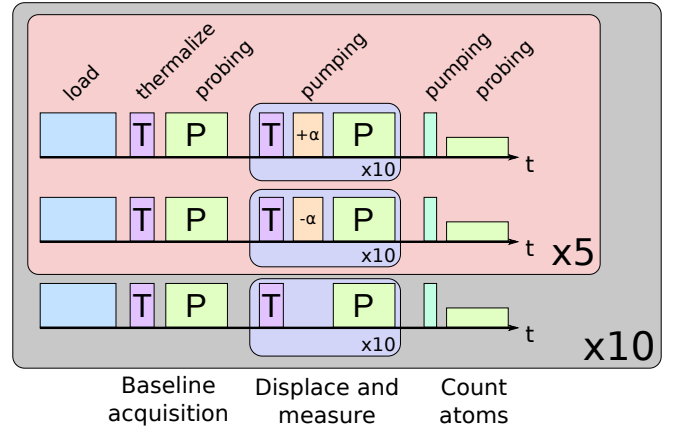}
\caption{\label{sequence}Experimental sequence:  
The experimental sequence divides into distinct tasks.
Baseline acquisition: prepare the thermal state and probe to measure the residual rotation.  Displace and measure (DM[$\alpha$]):  prepare the thermal state, displace by $\alpha$ and probe. Thanks to atom loss at each thermalization, the atom number is varied by repeating DM several times. Measure number of atoms $N_A$: by pumping the atoms into F=1, $m_F$=1 and probing we measure the number of atoms in the trap. To correct for drifts, a sequence without displacement (DM[0]) is performed every 11 runs. We perform the sequence varying the displacement to acquire a dataset of quantum-noise-limited measurements of $P_\alpha^{(\rm NG)}(S_y^{(\rm out)})$ for different $\alpha$. 
}%
\end{figure}

We generate Gaussian and non-Gaussian distributions with the following strategy:  we prepare a  ``thermal state'' (TS), an equal mixture of the $F=1, m_F=-1,0,1$ ground states, by repeated unpolarized optical pumping between the $F=1$ and $F=2$ hyperfine levels, finishing in  $F=1$ \cite{PhysRevLett.104.093602}.  By the central limit theorem, the TS of $10^6$ atoms is nearly Gaussian with $\expect{F_z} = 0$ and $\var(F_z) = \sigma^2 = 2N_A/3$. By optical pumping with pulses of circularly-polarized light we displace this to $\expect{F_z} = \alpha$, with negligible change in $\var(F_z)$ \cite{TothNJP2010}, to produce $P_{\alpha}(F_z) =  (\sigma \sqrt{2 \pi})^{-1} \exp[-(F_z-\alpha)^2/(2 \sigma^2)]$. By displacing different TS alternately to $\alpha_+$ and $\alpha_-$, we produce an equal statistical mixture of the two displaced states, $P_{\alpha}^{\rm (NG)}(F_z) = [P_{\alpha_+}(F_z) + P_{\alpha_-}(F_z)]/2$.
 With properly-chosen $\alpha_\pm$, $P_{\alpha}^{\rm (NG)}(F_z)$ closely approximates marginal distributions of mixtures of $n=0,1$ Fock states and $m=N,N-1$ symmetric Dicke states.  
The experimental sequence is shown in Fig. \ref{sequence}.

\newcommand{\SNR}{S}
\PRLsection{Detection, Analysis and Results}
For each preparation, 100  measurements of $F_z$ are made, with readings (i.e., estimated $F_z$ values by numerical integration of the measured signal)  $m_i = 2 S_y^{({\rm out},i)}/N_L^{(i)}$.  Because the measurement is non-destructive and shot noise limited, we can combine $\NPulses$ readings in a higher-sensitivity metapulse with reading $M\equiv \sum  m_i$ \cite{PhysRevLett.104.093602}.
 This has the distribution $P_{\alpha_\pm}(M) =   \exp[-(M- \alpha_\pm)^2/(2 \sigma_M^2)]/(\sigma_M \sqrt{2 \pi})$ where the variance $\sigma_M^2=\sigma_A^2 {N_A'}^2 \NPulses^2 + \sigma_R^2 $ includes atomic noise $\sigma_A^2 {N_A'}^2$  and readout noise, $\sigma_R^2 = \NPulses/N_L$ with $N_A'=N_A/N_A^{MAX}$ . The variance $\sigma_A^2$ is determined from the scaling of $\var(M)$ with $N_A$ and $\NPulses$, as in \cite{PhysRevLett.104.093602}. The readout noise can be varied over two orders of magnitude by appropriate choice of $\NPulses$. For one probe pulse and the maximum number of atoms we have $\sigma_R^2/\sigma_A^2=84.7$.

\begin{figure}
\includegraphics[width=1\linewidth]{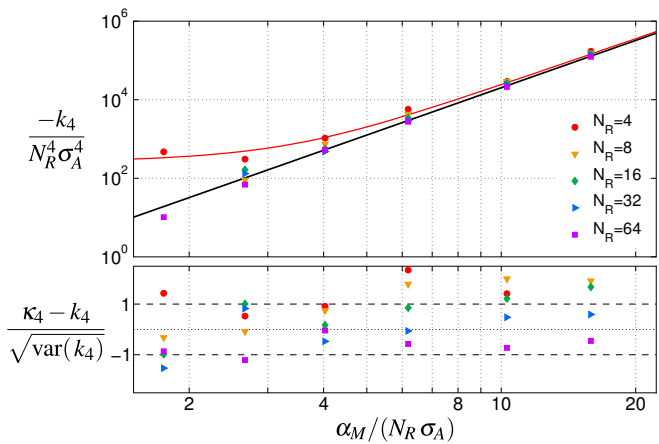}
\caption{\label{fig:exp:k4}
(color online) Measured and predicted $k_4$ with residuals for non-Gaussian distributions of different $\alpha$. Readout noise is varied by the choice of $\NPulses$. Data is normalized to $\NPulses$ and $\sigma_A$.
Top: Points show normalized $-k_4$ calculated from $N=100$ preparations of the ensemble with different $\alpha$ (horizontal axis), and  $\NPulses$ (colors).  Black line indicates expected $-\kappa_4$, red line (top) shows $-\kappa_4+\sqrt{\var(k_4)}$ calculated from the distribution parameters for the largest readout noise.
Bottom: normalized residuals  $(-k4+\kappa_4)/\sqrt{\var(k_4)}$. The normalization is done with the expected $\var(k_4)$ for each $\NPulses$.
 Measured $k_4$ agrees well with theory, in particular, measurement noise increases the observed variance, but not the expectation.}
\end{figure}

To produce a non-Gaussian distribution, we compose metapulses from $\NPulses$ samples drawn from displaced thermal state (DM[$\alpha_+$] or DM[$\alpha_-$]) preparations with equal probability, giving distribution $P_{\alpha}^{(\rm NG)}(M) = [ P_{\alpha_+}(M) + P_{\alpha_-}(M)]/2$. With $\alpha_M \equiv (\alpha_+-\alpha_-)/2$, the distribution has $\kappa_{2n+1}=0$, $\kappa_2 = \alpha_M^2 + \sigma_M^2$, $\kappa_4 = -2 \alpha_M^4$, $\kappa_6 = 16 \alpha_M^6$, $\kappa_8 = -272 \alpha_M^8$. Our ability to measure the non-Gaussianity is determined by $\expect{k_4} = \kappa_4$ and from Eq (\ref{eq:vark4})
\bea
\label{eq:vark4alpha}
\var(k_4) &=& 136 N \alpha_M^8 / N_{(1)} - 144 N^2 \alpha_M^4 (\alpha_M^2+\sigma_M^2)^2 / N_{(2)} \nnp 24 N^2(N+1) (\alpha_M^2+\sigma_M^2)^4 / N_{(3)}.
\eea
As shown in Fig. \ref{fig:exp:k4}, the experimentally obtained values agree well with theory, and confirm the independence from measurement noise.

\begin{figure}
\includegraphics[width=1\linewidth]{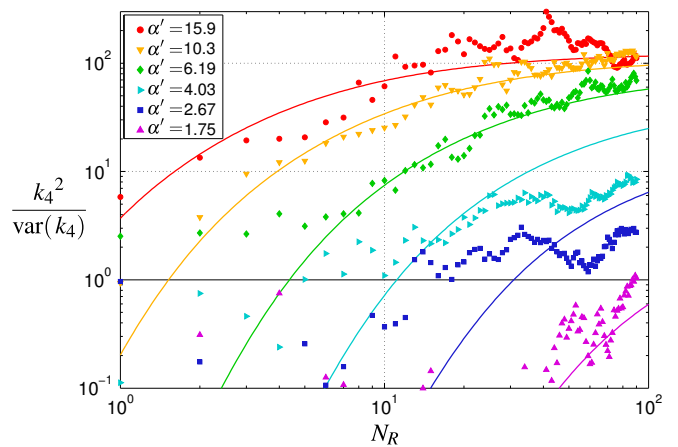}
\caption{\label{fig:exp:vark4}
(color online) Signal-to-noise in estimation of $\kappa_4$ versus readout noise for different $\alpha'=\alpha_M/(N_R \, \sigma_A)$. Points show measurement results, lines show theory. (details in the text)
}%
\end{figure}

The ``signal-to-noise ratio'' for $\kappa_4$, $\SNR = {\kappa_4}^2 / \var(k_4)$, is computed using Eq. (\ref{eq:vark4alpha}), $\kappa_4=-2 \alpha^4_M$, and experimental $\alpha_M$, $N_R$, $\sigma_R$, is shown as curves in Fig 4.  We can confirm this $\SNR$ experimentally by computing $S_N \equiv \expect{k_4}^2 / \var(k_4)$ using $k_4$ values derived from several realizations of the experiment, each sampling $P_\alpha^{NG}$ $N$ times.  In the limit of many realizations $S_N \rightarrow S$. We employ a bootstrapping technique:   From 100 samples of $P_{\alpha}^{(\rm NG)}(M)$ for given parameters $\alpha_M,\NPulses$ and $N_A$, we derive thirty-three $N=20$ realizations by random sampling without replacement, and compute $\expect{k_4}$ and $\var(k_4)$ on the realizations. As shown in Fig. \ref{fig:exp:vark4}, good agreement with theory is observed.

\PRLsection{Optimum estimation of non-Gaussian distributions}
Finally, we note that in scenarios where measurements are expensive relative to state preparation (as might be the case for QND measurements of optical fields or for testing the successful storage of a single photon in a quantum memory), optimal use of measurement resources (e.g. measurement time) avoids both too few preparations and too few probings.

We consider a scenario of practical interest for quantum networking:  a heralded single-photon state is produced and stored in an atomic ensemble quantum memory.  Assuming the ensemble is initially polarized in the $\hat{X}$ direction, the storage process maps the quadrature components $X,P$ onto the corresponding atomic spin operators $X_{A},P_{A} \propto F_z, -F_y$, respectively.   QND measurements of $F_z$  are used to estimate $X_A$, and thus the non-Gaussianity of the stored single photon.   Due to imperfect storage, this will have the distribution of a mixture of $n=0$ and $n=1$ Fock states: $\rho = (1-p) \left|0\rangle \langle 0\right| +p \left|1\rangle \langle 1\right|$. For a quadrature $X$, we have the following probability distribution $P_{p}(X)=\exp[{-{x^2}/(2 \sigma_0^2)}] \left(p x^2/\sigma_0^2+1-p\right)/\left(\sqrt{2 \pi } \sigma_0\right)$, where $\sigma_0$ is the width of the $n=0$ state. 

Taking in account the readout noise $\sigma_R^2$, the cumulants are $\kappa_{\rm odd}=0$, $\kappa_2 = (2 p+1) \sigma _H^2 + \sigma_R^2$, $\kappa_4 = -12 p^2 \sigma _H^4$, $\kappa_6 = 240 p^3 \sigma _H^6$, $\kappa_8 =-10080 p^4 \sigma _H^8$, where the readout noise $\sigma_R^2$ is included as above.
 Here $\kappa_4$ is directly related to the classicality of the state, since $p>0.5$ implies a negative Wigner distribution \cite{PhysRevLett.87.050402}.

\begin{figure}
\includegraphics[width=0.8\linewidth]{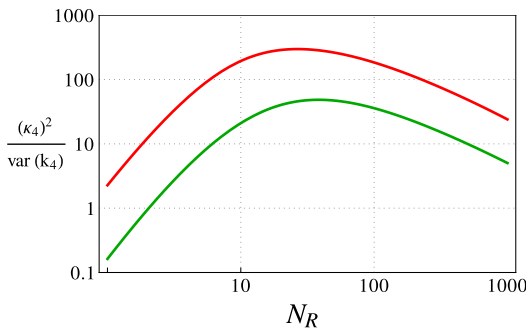}
\caption{(color online) Signal-to noise-ratio $\SNR$ versus $\NPulses$ for a fixed probe number $N_M \NPulses = \num{1e5}$ for the probability distribution associated with Fock state mixture described in the text with a normalized n=0 width $\sigma_0=1$.
Red curve (top): p=1.  Green curve (bottom): p= 0.5
with SNL measurement: $\sigma_R=\sqrt{20/\NPulses}$. 
 \label{fig:th_opt_ats}}
\end{figure}

For a fixed total number of measurement resources $N_M \NPulses$, an optimal distribution of resources per measurement $\NPulses$ exists as shown in Fig. \ref{fig:th_opt_ats}. With increasing $\NPulses$, the signal-to-noise first increases due to the improvement of the measurement precision. Then, once the increased measurement precision no longer gives extra information about $k_4$, the precision decreases due to reduced statistics because of the limited total number of probes.
For a large total number of measurements, we can derive a simplified expression of this optimum. 
We derive asymptotic expressions of $S$: $S_L$ ($S_H$) for $\sigma_R \ll \sigma_0$ ($\sigma_R \gg \sigma_0$). The optimal $\NPulses$ is found by solving $S_L = S_R$ giving ${\sigma_R}^8 \approx \sigma _0^8 (1+8p-12p^2 +48p^3 - 24p^4)$. For this optimal $\sigma_R$, the measurement noise is in the same order of magnitude as the characteristic width of the non-Gaussian distribution.

\PRLsection{Conclusion}
The cumulant-based methods described here should be very attractive for experiments with non-Gaussian states of material systems such as atomic ensembles and nano-resonators, for which the state preparation time is intrinsically longer, and for which measurement noise is a greater challenge than in optical systems. Cumulant-based estimation is simultaneously efficient, requiring few preparations and measurements, accommodates measurement noise in a natural way, and facilitates statistically-meaningful tests, e.g., of  non-classicality.  Experimental tests with a cold atomic ensemble demonstrate the method in a system highly suitable for quantum networking, while the theory applies equally to other quantum systems of current interest.

\begin{acknowledgments} 
\PRLsection{Acknowledgements} This project has been funded by The Spanish Ministry of Science and Innovation under the ILUMA project (Ref. FIS2008-01051), the Consider-Ingenio 2010 Project "QOIT" and the Marie-Curie RTN EMALI.
\end{acknowledgments}

\bibliographystyle{apsrev4-1}

\begin{thebibliography}{27}%
\makeatletter
\providecommand \@ifxundefined [1]{%
 \@ifx{#1\undefined}
}%
\providecommand \@ifnum [1]{%
 \ifnum #1\expandafter \@firstoftwo
 \else \expandafter \@secondoftwo
 \fi
}%
\providecommand \@ifx [1]{%
 \ifx #1\expandafter \@firstoftwo
 \else \expandafter \@secondoftwo
 \fi
}%
\providecommand \natexlab [1]{#1}%
\providecommand \enquote  [1]{``#1''}%
\providecommand \bibnamefont  [1]{#1}%
\providecommand \bibfnamefont [1]{#1}%
\providecommand \citenamefont [1]{#1}%
\providecommand \href@noop [0]{\@secondoftwo}%
\providecommand \href [0]{\begingroup \@sanitize@url \@href}%
\providecommand \@href[1]{\@@startlink{#1}\@@href}%
\providecommand \@@href[1]{\endgroup#1\@@endlink}%
\providecommand \@sanitize@url [0]{\catcode `\\12\catcode `\$12\catcode
  `\&12\catcode `\#12\catcode `\^12\catcode `\_12\catcode `\%12\relax}%
\providecommand \@@startlink[1]{}%
\providecommand \@@endlink[0]{}%
\providecommand \url  [0]{\begingroup\@sanitize@url \@url }%
\providecommand \@url [1]{\endgroup\@href {#1}{\urlprefix }}%
\providecommand \urlprefix  [0]{URL }%
\providecommand \Eprint [0]{\href }%
\providecommand \doibase [0]{http://dx.doi.org/}%
\providecommand \selectlanguage [0]{\@gobble}%
\providecommand \bibinfo  [0]{\@secondoftwo}%
\providecommand \bibfield  [0]{\@secondoftwo}%
\providecommand \translation [1]{[#1]}%
\providecommand \BibitemOpen [0]{}%
\providecommand \bibitemStop [0]{}%
\providecommand \bibitemNoStop [0]{.\EOS\space}%
\providecommand \EOS [0]{\spacefactor3000\relax}%
\providecommand \BibitemShut  [1]{\csname bibitem#1\endcsname}%
\let\auto@bib@innerbib\@empty
\bibitem [{\citenamefont {Ralph}\ \emph {et~al.}(2003)\citenamefont {Ralph},
  \citenamefont {Gilchrist}, \citenamefont {Milburn}, \citenamefont {Munro},\
  and\ \citenamefont {Glancy}}]{PhysRevA.68.042319}%
  \BibitemOpen
  \bibfield  {author} {\bibinfo {author} {\bibfnamefont {T.~C.}\ \bibnamefont
  {Ralph}}, \bibinfo {author} {\bibfnamefont {A.}~\bibnamefont {Gilchrist}},
  \bibinfo {author} {\bibfnamefont {G.~J.}\ \bibnamefont {Milburn}}, \bibinfo
  {author} {\bibfnamefont {W.~J.}\ \bibnamefont {Munro}}, \ and\ \bibinfo
  {author} {\bibfnamefont {S.}~\bibnamefont {Glancy}},\ }\href {\doibase
  10.1103/PhysRevA.68.042319} {\bibfield  {journal} {\bibinfo  {journal} {Phys.
  Rev. A}\ }\textbf {\bibinfo {volume} {68}},\ \bibinfo {pages} {042319}
  (\bibinfo {year} {2003})}\BibitemShut {NoStop}%
\bibitem [{\citenamefont {Lloyd}\ and\ \citenamefont
  {Braunstein}(1999)}]{PhysRevLett.82.1784}%
  \BibitemOpen
  \bibfield  {author} {\bibinfo {author} {\bibfnamefont {S.}~\bibnamefont
  {Lloyd}}\ and\ \bibinfo {author} {\bibfnamefont {S.~L.}\ \bibnamefont
  {Braunstein}},\ }\href {\doibase 10.1103/PhysRevLett.82.1784} {\bibfield
  {journal} {\bibinfo  {journal} {Phys. Rev. Lett.}\ }\textbf {\bibinfo
  {volume} {82}},\ \bibinfo {pages} {1784} (\bibinfo {year}
  {1999})}\BibitemShut {NoStop}%
\bibitem [{\citenamefont {Dell'Anno}\ \emph {et~al.}(2007)\citenamefont
  {Dell'Anno}, \citenamefont {De~Siena}, \citenamefont {Albano},\ and\
  \citenamefont {Illuminati}}]{PhysRevA.76.022301}%
  \BibitemOpen
  \bibfield  {author} {\bibinfo {author} {\bibfnamefont {F.}~\bibnamefont
  {Dell'Anno}}, \bibinfo {author} {\bibfnamefont {S.}~\bibnamefont {De~Siena}},
  \bibinfo {author} {\bibfnamefont {L.}~\bibnamefont {Albano}}, \ and\ \bibinfo
  {author} {\bibfnamefont {F.}~\bibnamefont {Illuminati}},\ }\href {\doibase
  10.1103/PhysRevA.76.022301} {\bibfield  {journal} {\bibinfo  {journal} {Phys.
  Rev. A}\ }\textbf {\bibinfo {volume} {76}},\ \bibinfo {pages} {022301}
  (\bibinfo {year} {2007})}\BibitemShut {NoStop}%
\bibitem [{\citenamefont {Eisert}\ \emph {et~al.}(2002)\citenamefont {Eisert},
  \citenamefont {Scheel},\ and\ \citenamefont
  {Plenio}}]{PhysRevLett.89.137903}%
  \BibitemOpen
  \bibfield  {author} {\bibinfo {author} {\bibfnamefont {J.}~\bibnamefont
  {Eisert}}, \bibinfo {author} {\bibfnamefont {S.}~\bibnamefont {Scheel}}, \
  and\ \bibinfo {author} {\bibfnamefont {M.~B.}\ \bibnamefont {Plenio}},\
  }\href {\doibase 10.1103/PhysRevLett.89.137903} {\bibfield  {journal}
  {\bibinfo  {journal} {Phys. Rev. Lett.}\ }\textbf {\bibinfo {volume} {89}},\
  \bibinfo {pages} {137903} (\bibinfo {year} {2002})}\BibitemShut {NoStop}%
\bibitem [{\citenamefont {Giedke}\ and\ \citenamefont
  {Ignacio~Cirac}(2002)}]{PhysRevA.66.032316}%
  \BibitemOpen
  \bibfield  {author} {\bibinfo {author} {\bibfnamefont {G.}~\bibnamefont
  {Giedke}}\ and\ \bibinfo {author} {\bibfnamefont {J.}~\bibnamefont
  {Ignacio~Cirac}},\ }\href {\doibase 10.1103/PhysRevA.66.032316} {\bibfield
  {journal} {\bibinfo  {journal} {Phys. Rev. A}\ }\textbf {\bibinfo {volume}
  {66}},\ \bibinfo {pages} {032316} (\bibinfo {year} {2002})}\BibitemShut
  {NoStop}%
\bibitem [{\citenamefont {Neergaard-Nielsen}\ \emph {et~al.}(2006)\citenamefont
  {Neergaard-Nielsen}, \citenamefont {Nielsen}, \citenamefont {Hettich},
  \citenamefont {M\o{}lmer},\ and\ \citenamefont
  {Polzik}}]{PhysRevLett.97.083604}%
  \BibitemOpen
  \bibfield  {author} {\bibinfo {author} {\bibfnamefont {J.~S.}\ \bibnamefont
  {Neergaard-Nielsen}}, \bibinfo {author} {\bibfnamefont {B.~M.}\ \bibnamefont
  {Nielsen}}, \bibinfo {author} {\bibfnamefont {C.}~\bibnamefont {Hettich}},
  \bibinfo {author} {\bibfnamefont {K.}~\bibnamefont {M\o{}lmer}}, \ and\
  \bibinfo {author} {\bibfnamefont {E.~S.}\ \bibnamefont {Polzik}},\ }\href
  {\doibase 10.1103/PhysRevLett.97.083604} {\bibfield  {journal} {\bibinfo
  {journal} {Phys. Rev. Lett.}\ }\textbf {\bibinfo {volume} {97}},\ \bibinfo
  {pages} {083604} (\bibinfo {year} {2006})}\BibitemShut {NoStop}%
\bibitem [{\citenamefont {Ourjoumtsev}\ \emph {et~al.}(2007)\citenamefont
  {Ourjoumtsev}, \citenamefont {Jeong}, \citenamefont {Tualle-Brouri},\ and\
  \citenamefont {Grangier}}]{Ourjoumtsev2007}%
  \BibitemOpen
  \bibfield  {author} {\bibinfo {author} {\bibfnamefont {A.}~\bibnamefont
  {Ourjoumtsev}}, \bibinfo {author} {\bibfnamefont {H.}~\bibnamefont {Jeong}},
  \bibinfo {author} {\bibfnamefont {R.}~\bibnamefont {Tualle-Brouri}}, \ and\
  \bibinfo {author} {\bibfnamefont {P.}~\bibnamefont {Grangier}},\ }\href
  {http://dx.doi.org/10.1038/nature06054} {\bibfield  {journal} {\bibinfo
  {journal} {Nature}\ }\textbf {\bibinfo {volume} {448}},\ \bibinfo {pages}
  {784} (\bibinfo {year} {2007})}\BibitemShut {NoStop}%
\bibitem [{\citenamefont {Wakui}\ \emph {et~al.}(2007)\citenamefont {Wakui},
  \citenamefont {Takahashi}, \citenamefont {Furusawa},\ and\ \citenamefont
  {Sasaki}}]{Wakui:07}%
  \BibitemOpen
  \bibfield  {author} {\bibinfo {author} {\bibfnamefont {K.}~\bibnamefont
  {Wakui}}, \bibinfo {author} {\bibfnamefont {H.}~\bibnamefont {Takahashi}},
  \bibinfo {author} {\bibfnamefont {A.}~\bibnamefont {Furusawa}}, \ and\
  \bibinfo {author} {\bibfnamefont {M.}~\bibnamefont {Sasaki}},\ }\href
  {\doibase 10.1364/OE.15.003568} {\bibfield  {journal} {\bibinfo  {journal}
  {Opt. Express}\ }\textbf {\bibinfo {volume} {15}},\ \bibinfo {pages} {3568}
  (\bibinfo {year} {2007})}\BibitemShut {NoStop}%
\bibitem [{\citenamefont {Takahashi}\ \emph {et~al.}(2008)\citenamefont
  {Takahashi}, \citenamefont {Wakui}, \citenamefont {Suzuki}, \citenamefont
  {Takeoka}, \citenamefont {Hayasaka}, \citenamefont {Furusawa},\ and\
  \citenamefont {Sasaki}}]{PhysRevLett.101.233605}%
  \BibitemOpen
  \bibfield  {author} {\bibinfo {author} {\bibfnamefont {H.}~\bibnamefont
  {Takahashi}}, \bibinfo {author} {\bibfnamefont {K.}~\bibnamefont {Wakui}},
  \bibinfo {author} {\bibfnamefont {S.}~\bibnamefont {Suzuki}}, \bibinfo
  {author} {\bibfnamefont {M.}~\bibnamefont {Takeoka}}, \bibinfo {author}
  {\bibfnamefont {K.}~\bibnamefont {Hayasaka}}, \bibinfo {author}
  {\bibfnamefont {A.}~\bibnamefont {Furusawa}}, \ and\ \bibinfo {author}
  {\bibfnamefont {M.}~\bibnamefont {Sasaki}},\ }\href {\doibase
  10.1103/PhysRevLett.101.233605} {\bibfield  {journal} {\bibinfo  {journal}
  {Phys. Rev. Lett.}\ }\textbf {\bibinfo {volume} {101}},\ \bibinfo {pages}
  {233605} (\bibinfo {year} {2008})}\BibitemShut {NoStop}%
\bibitem [{\citenamefont {Je\ifmmode~\check{z}\else \v{z}\fi{}ek}\ \emph
  {et~al.}(2011)\citenamefont {Je\ifmmode~\check{z}\else \v{z}\fi{}ek},
  \citenamefont {Straka}, \citenamefont {Mi\ifmmode~\check{c}\else
  \v{c}\fi{}uda}, \citenamefont {Du\ifmmode~\check{s}\else \v{s}\fi{}ek},
  \citenamefont {Fiur\'a\ifmmode~\check{s}\else \v{s}\fi{}ek},\ and\
  \citenamefont {Filip}}]{PhysRevLett.107.213602}%
  \BibitemOpen
  \bibfield  {author} {\bibinfo {author} {\bibfnamefont {M.}~\bibnamefont
  {Je\ifmmode~\check{z}\else \v{z}\fi{}ek}}, \bibinfo {author} {\bibfnamefont
  {I.}~\bibnamefont {Straka}}, \bibinfo {author} {\bibfnamefont
  {M.}~\bibnamefont {Mi\ifmmode~\check{c}\else \v{c}\fi{}uda}}, \bibinfo
  {author} {\bibfnamefont {M.}~\bibnamefont {Du\ifmmode~\check{s}\else
  \v{s}\fi{}ek}}, \bibinfo {author} {\bibfnamefont {J.}~\bibnamefont
  {Fiur\'a\ifmmode~\check{s}\else \v{s}\fi{}ek}}, \ and\ \bibinfo {author}
  {\bibfnamefont {R.}~\bibnamefont {Filip}},\ }\href {\doibase
  10.1103/PhysRevLett.107.213602} {\bibfield  {journal} {\bibinfo  {journal}
  {Phys. Rev. Lett.}\ }\textbf {\bibinfo {volume} {107}},\ \bibinfo {pages}
  {213602} (\bibinfo {year} {2011})}\BibitemShut {NoStop}%
\bibitem [{\citenamefont {Massar}\ and\ \citenamefont
  {Polzik}(2003)}]{PhysRevLett.91.060401}%
  \BibitemOpen
  \bibfield  {author} {\bibinfo {author} {\bibfnamefont {S.}~\bibnamefont
  {Massar}}\ and\ \bibinfo {author} {\bibfnamefont {E.~S.}\ \bibnamefont
  {Polzik}},\ }\href {\doibase 10.1103/PhysRevLett.91.060401} {\bibfield
  {journal} {\bibinfo  {journal} {Phys. Rev. Lett.}\ }\textbf {\bibinfo
  {volume} {91}},\ \bibinfo {pages} {060401} (\bibinfo {year}
  {2003})}\BibitemShut {NoStop}%
\bibitem [{\citenamefont {Nielsen}\ \emph {et~al.}(2009)\citenamefont
  {Nielsen}, \citenamefont {Poulsen}, \citenamefont {Negretti},\ and\
  \citenamefont {M\o{}lmer}}]{PhysRevA.79.023841}%
  \BibitemOpen
  \bibfield  {author} {\bibinfo {author} {\bibfnamefont {A.~E.~B.}\
  \bibnamefont {Nielsen}}, \bibinfo {author} {\bibfnamefont {U.~V.}\
  \bibnamefont {Poulsen}}, \bibinfo {author} {\bibfnamefont {A.}~\bibnamefont
  {Negretti}}, \ and\ \bibinfo {author} {\bibfnamefont {K.}~\bibnamefont
  {M\o{}lmer}},\ }\href {\doibase 10.1103/PhysRevA.79.023841} {\bibfield
  {journal} {\bibinfo  {journal} {Phys. Rev. A}\ }\textbf {\bibinfo {volume}
  {79}},\ \bibinfo {pages} {023841} (\bibinfo {year} {2009})}\BibitemShut
  {NoStop}%
\bibitem [{\citenamefont {Lemr}\ and\ \citenamefont
  {Fiur\'a\ifmmode~\check{s}\else \v{s}\fi{}ek}(2009)}]{PhysRevA.79.043808}%
  \BibitemOpen
  \bibfield  {author} {\bibinfo {author} {\bibfnamefont {K.}~\bibnamefont
  {Lemr}}\ and\ \bibinfo {author} {\bibfnamefont {J.}~\bibnamefont
  {Fiur\'a\ifmmode~\check{s}\else \v{s}\fi{}ek}},\ }\href {\doibase
  10.1103/PhysRevA.79.043808} {\bibfield  {journal} {\bibinfo  {journal} {Phys.
  Rev. A}\ }\textbf {\bibinfo {volume} {79}},\ \bibinfo {pages} {043808}
  (\bibinfo {year} {2009})}\BibitemShut {NoStop}%
\bibitem [{\citenamefont {Mazets}\ \emph {et~al.}(2008)\citenamefont {Mazets},
  \citenamefont {Kurizki}, \citenamefont {Oberthaler},\ and\ \citenamefont
  {Schmiedmayer}}]{0295-5075-83-6-60004}%
  \BibitemOpen
  \bibfield  {author} {\bibinfo {author} {\bibfnamefont {I.~E.}\ \bibnamefont
  {Mazets}}, \bibinfo {author} {\bibfnamefont {G.}~\bibnamefont {Kurizki}},
  \bibinfo {author} {\bibfnamefont {M.~K.}\ \bibnamefont {Oberthaler}}, \ and\
  \bibinfo {author} {\bibfnamefont {J.}~\bibnamefont {Schmiedmayer}},\ }\href
  {http://stacks.iop.org/0295-5075/83/i=6/a=60004} {\bibfield  {journal}
  {\bibinfo  {journal} {EPL (Europhysics Letters)}\ }\textbf {\bibinfo {volume}
  {83}},\ \bibinfo {pages} {60004} (\bibinfo {year} {2008})}\BibitemShut
  {NoStop}%
\bibitem [{\citenamefont {Wenger}\ \emph {et~al.}(2004)\citenamefont {Wenger},
  \citenamefont {Tualle-Brouri},\ and\ \citenamefont
  {Grangier}}]{PhysRevLett.92.153601}%
  \BibitemOpen
  \bibfield  {author} {\bibinfo {author} {\bibfnamefont {J.}~\bibnamefont
  {Wenger}}, \bibinfo {author} {\bibfnamefont {R.}~\bibnamefont
  {Tualle-Brouri}}, \ and\ \bibinfo {author} {\bibfnamefont {P.}~\bibnamefont
  {Grangier}},\ }\href {\doibase 10.1103/PhysRevLett.92.153601} {\bibfield
  {journal} {\bibinfo  {journal} {Phys. Rev. Lett.}\ }\textbf {\bibinfo
  {volume} {92}},\ \bibinfo {pages} {153601} (\bibinfo {year}
  {2004})}\BibitemShut {NoStop}%
\bibitem [{\citenamefont {Bednorz}\ and\ \citenamefont
  {Belzig}(2011)}]{PhysRevA.83.052113}%
  \BibitemOpen
  \bibfield  {author} {\bibinfo {author} {\bibfnamefont {A.}~\bibnamefont
  {Bednorz}}\ and\ \bibinfo {author} {\bibfnamefont {W.}~\bibnamefont
  {Belzig}},\ }\href {\doibase 10.1103/PhysRevA.83.052113} {\bibfield
  {journal} {\bibinfo  {journal} {Phys. Rev. A}\ }\textbf {\bibinfo {volume}
  {83}},\ \bibinfo {pages} {052113} (\bibinfo {year} {2011})}\BibitemShut
  {NoStop}%
\bibitem [{\citenamefont {Shchukin}\ \emph {et~al.}(2005)\citenamefont
  {Shchukin}, \citenamefont {Richter},\ and\ \citenamefont
  {Vogel}}]{PhysRevA.71.011802}%
  \BibitemOpen
  \bibfield  {author} {\bibinfo {author} {\bibfnamefont {E.}~\bibnamefont
  {Shchukin}}, \bibinfo {author} {\bibfnamefont {T.}~\bibnamefont {Richter}}, \
  and\ \bibinfo {author} {\bibfnamefont {W.}~\bibnamefont {Vogel}},\ }\href
  {\doibase 10.1103/PhysRevA.71.011802} {\bibfield  {journal} {\bibinfo
  {journal} {Phys. Rev. A}\ }\textbf {\bibinfo {volume} {71}},\ \bibinfo
  {pages} {011802} (\bibinfo {year} {2005})}\BibitemShut {NoStop}%
\bibitem [{\citenamefont {Eran~Kot}\ and\ \citenamefont
  {S{\o}rensen}(2011)}]{arxiv1110.3060}%
  \BibitemOpen
  \bibfield  {author} {\bibinfo {author} {\bibfnamefont {E.~S.~P.}\
  \bibnamefont {Eran~Kot}, \bibfnamefont {Niels Gr{\o}nbech-Jensen}}\ and\
  \bibinfo {author} {\bibfnamefont {A.~S.}\ \bibnamefont {S{\o}rensen}},\
  }\href@noop {} {\bibfield  {journal} {\bibinfo  {journal} {arXiv}\ }
  (\bibinfo {year} {2011})},\ \Eprint
  {http://arxiv.org/abs/arXiv:quant-ph/1110.3060} {arXiv:quant-ph/1110.3060}
  \BibitemShut {NoStop}%
\bibitem [{\citenamefont {Lvovsky}\ \emph {et~al.}(2001)\citenamefont
  {Lvovsky}, \citenamefont {Hansen}, \citenamefont {Aichele}, \citenamefont
  {Benson}, \citenamefont {Mlynek},\ and\ \citenamefont
  {Schiller}}]{PhysRevLett.87.050402}%
  \BibitemOpen
  \bibfield  {author} {\bibinfo {author} {\bibfnamefont {A.~I.}\ \bibnamefont
  {Lvovsky}}, \bibinfo {author} {\bibfnamefont {H.}~\bibnamefont {Hansen}},
  \bibinfo {author} {\bibfnamefont {T.}~\bibnamefont {Aichele}}, \bibinfo
  {author} {\bibfnamefont {O.}~\bibnamefont {Benson}}, \bibinfo {author}
  {\bibfnamefont {J.}~\bibnamefont {Mlynek}}, \ and\ \bibinfo {author}
  {\bibfnamefont {S.}~\bibnamefont {Schiller}},\ }\href {\doibase
  10.1103/PhysRevLett.87.050402} {\bibfield  {journal} {\bibinfo  {journal}
  {Phys. Rev. Lett.}\ }\textbf {\bibinfo {volume} {87}},\ \bibinfo {pages}
  {050402} (\bibinfo {year} {2001})}\BibitemShut {NoStop}%
\bibitem [{\citenamefont {Appel}\ \emph {et~al.}(2009)\citenamefont {Appel},
  \citenamefont {Windpassinger}, \citenamefont {Oblak}, \citenamefont {Hoff},
  \citenamefont {Kj{\ae}rgaard},\ and\ \citenamefont {Polzik}}]{Appel07072009}%
  \BibitemOpen
  \bibfield  {author} {\bibinfo {author} {\bibfnamefont {J.}~\bibnamefont
  {Appel}}, \bibinfo {author} {\bibfnamefont {P.~J.}\ \bibnamefont
  {Windpassinger}}, \bibinfo {author} {\bibfnamefont {D.}~\bibnamefont
  {Oblak}}, \bibinfo {author} {\bibfnamefont {U.~B.}\ \bibnamefont {Hoff}},
  \bibinfo {author} {\bibfnamefont {N.}~\bibnamefont {Kj{\ae}rgaard}}, \ and\
  \bibinfo {author} {\bibfnamefont {E.~S.}\ \bibnamefont {Polzik}},\ }\href
  {\doibase 10.1073/pnas.0901550106} {\bibfield  {journal} {\bibinfo  {journal}
  {Proceedings of the National Academy of Sciences}\ }\textbf {\bibinfo
  {volume} {106}},\ \bibinfo {pages} {10960} (\bibinfo {year}
  {2009})}\BibitemShut {NoStop}%
\bibitem [{\citenamefont {Koschorreck}\ \emph
  {et~al.}(2010{\natexlab{a}})\citenamefont {Koschorreck}, \citenamefont
  {Napolitano}, \citenamefont {Dubost},\ and\ \citenamefont
  {Mitchell}}]{PhysRevLett.105.093602}%
  \BibitemOpen
  \bibfield  {author} {\bibinfo {author} {\bibfnamefont {M.}~\bibnamefont
  {Koschorreck}}, \bibinfo {author} {\bibfnamefont {M.}~\bibnamefont
  {Napolitano}}, \bibinfo {author} {\bibfnamefont {B.}~\bibnamefont {Dubost}},
  \ and\ \bibinfo {author} {\bibfnamefont {M.~W.}\ \bibnamefont {Mitchell}},\
  }\href {\doibase 10.1103/PhysRevLett.105.093602} {\bibfield  {journal}
  {\bibinfo  {journal} {Phys. Rev. Lett.}\ }\textbf {\bibinfo {volume} {105}},\
  \bibinfo {pages} {093602} (\bibinfo {year} {2010}{\natexlab{a}})}\BibitemShut
  {NoStop}%
\bibitem [{\citenamefont {Hertzberg}\ \emph {et~al.}(2010)\citenamefont
  {Hertzberg}, \citenamefont {Rocheleau}, \citenamefont {Ndukum}, \citenamefont
  {Savva}, \citenamefont {Clerk},\ and\ \citenamefont
  {Schwab}}]{HertzbergNP2010}%
  \BibitemOpen
  \bibfield  {author} {\bibinfo {author} {\bibfnamefont {J.~B.}\ \bibnamefont
  {Hertzberg}}, \bibinfo {author} {\bibfnamefont {T.}~\bibnamefont
  {Rocheleau}}, \bibinfo {author} {\bibfnamefont {T.}~\bibnamefont {Ndukum}},
  \bibinfo {author} {\bibfnamefont {M.}~\bibnamefont {Savva}}, \bibinfo
  {author} {\bibfnamefont {A.~A.}\ \bibnamefont {Clerk}}, \ and\ \bibinfo
  {author} {\bibfnamefont {K.~C.}\ \bibnamefont {Schwab}},\ }\href
  {http://dx.doi.org/10.1038/nphys1479} {\bibfield  {journal} {\bibinfo
  {journal} {Nat Phys}\ }\textbf {\bibinfo {volume} {6}},\ \bibinfo {pages}
  {213} (\bibinfo {year} {2010})}\BibitemShut {NoStop}%
\bibitem [{\citenamefont {Kendall}\ and\ \citenamefont
  {Stuart}(1958)}]{AdvTheoStat1958}%
  \BibitemOpen
  \bibfield  {author} {\bibinfo {author} {\bibfnamefont {M.~G.}\ \bibnamefont
  {Kendall}}\ and\ \bibinfo {author} {\bibfnamefont {A.}~\bibnamefont
  {Stuart}},\ }\href@noop {} {\emph {\bibinfo {title} {The advanced theory of
  statistics}}}\ (\bibinfo  {publisher} {C. Griffin, London},\ \bibinfo {year}
  {1958})\BibitemShut {NoStop}%
\bibitem [{\citenamefont {Polzik}\ and\ \citenamefont
  {M\"uller}()}]{JHMPersComm}%
  \BibitemOpen
  \bibfield  {author} {\bibinfo {author} {\bibfnamefont {E.~S.}\ \bibnamefont
  {Polzik}}\ and\ \bibinfo {author} {\bibfnamefont {J.~H.}\ \bibnamefont
  {M\"uller}},\ }\href@noop {} {}\bibinfo {howpublished} {pers.
  comm.}\BibitemShut {Stop}%
\bibitem [{\citenamefont {Kubasik}\ \emph {et~al.}(2009)\citenamefont
  {Kubasik}, \citenamefont {Koschorreck}, \citenamefont {Napolitano},
  \citenamefont {de~Echaniz}, \citenamefont {Crepaz}, \citenamefont {Eschner},
  \citenamefont {Polzik},\ and\ \citenamefont {Mitchell}}]{PhysRevA.79.043815}%
  \BibitemOpen
  \bibfield  {author} {\bibinfo {author} {\bibfnamefont {M.}~\bibnamefont
  {Kubasik}}, \bibinfo {author} {\bibfnamefont {M.}~\bibnamefont
  {Koschorreck}}, \bibinfo {author} {\bibfnamefont {M.}~\bibnamefont
  {Napolitano}}, \bibinfo {author} {\bibfnamefont {S.~R.}\ \bibnamefont
  {de~Echaniz}}, \bibinfo {author} {\bibfnamefont {H.}~\bibnamefont {Crepaz}},
  \bibinfo {author} {\bibfnamefont {J.}~\bibnamefont {Eschner}}, \bibinfo
  {author} {\bibfnamefont {E.~S.}\ \bibnamefont {Polzik}}, \ and\ \bibinfo
  {author} {\bibfnamefont {M.~W.}\ \bibnamefont {Mitchell}},\ }\href {\doibase
  10.1103/PhysRevA.79.043815} {\bibfield  {journal} {\bibinfo  {journal} {Phys.
  Rev. A}\ }\textbf {\bibinfo {volume} {79}},\ \bibinfo {pages} {043815}
  (\bibinfo {year} {2009})}\BibitemShut {NoStop}%
\bibitem [{\citenamefont {Koschorreck}\ \emph
  {et~al.}(2010{\natexlab{b}})\citenamefont {Koschorreck}, \citenamefont
  {Napolitano}, \citenamefont {Dubost},\ and\ \citenamefont
  {Mitchell}}]{PhysRevLett.104.093602}%
  \BibitemOpen
  \bibfield  {author} {\bibinfo {author} {\bibfnamefont {M.}~\bibnamefont
  {Koschorreck}}, \bibinfo {author} {\bibfnamefont {M.}~\bibnamefont
  {Napolitano}}, \bibinfo {author} {\bibfnamefont {B.}~\bibnamefont {Dubost}},
  \ and\ \bibinfo {author} {\bibfnamefont {M.~W.}\ \bibnamefont {Mitchell}},\
  }\href {\doibase 10.1103/PhysRevLett.104.093602} {\bibfield  {journal}
  {\bibinfo  {journal} {Phys. Rev. Lett.}\ }\textbf {\bibinfo {volume} {104}},\
  \bibinfo {pages} {093602} (\bibinfo {year} {2010}{\natexlab{b}})}\BibitemShut
  {NoStop}%
\bibitem [{\citenamefont {T{\'o}th}\ and\ \citenamefont
  {Mitchell}(2010)}]{TothNJP2010}%
  \BibitemOpen
  \bibfield  {author} {\bibinfo {author} {\bibfnamefont {G.}~\bibnamefont
  {T{\'o}th}}\ and\ \bibinfo {author} {\bibfnamefont {M.~W.}\ \bibnamefont
  {Mitchell}},\ }\href {http://stacks.iop.org/1367-2630/12/i=5/a=053007}
  {\bibfield  {journal} {\bibinfo  {journal} {New Journal of Physics}\ }\textbf
  {\bibinfo {volume} {12}},\ \bibinfo {pages} {053007} (\bibinfo {year}
  {2010})}\BibitemShut {NoStop}%
\end{thebibliography}

%

\end{document}